# Highly enhanced field emission current density of copper oxide coated vertically aligned carbon nanotubes: Role of interface and electronic structure

M. Sreekanth[1], S. Ghosh[1], and P. Srivastava[1*]

[1]Nanostech Laboratory, Department of Physics, Indian Institute of Technology Delhi, Hauz Khas, New Delhi-110016, India

**Abstract**

We report the field emission (FE) properties of Cu coated vertically aligned carbon nanotubes (VACNTs) before and after oxidation. The current density was found to be the maximum (20 mA/cm$^2$) for 3 nm thick Cu coated VACNTs after oxidation. The variation in conventionally monitored parameters like work function and field enhancement factor does not explain the experimentally determined FE current density. A critical analysis of the electronic structure reveals the importance of presence of Cu$_2$O at the interface of CuO and VACNTs, which in turn controls the current density of these films. The highly enhanced FE current density of 3 nm Cu coated VACNTs after oxidation suggests its potential as a next generation electron source in vacuum microelectronic devices.

[*]**Corresponding author e-mail:** pankajs@physics.iitd.ernet.in (P. Srivastava)

# 1. Introduction

Carbon nanotubes [1] (CNTs) are one of the most important nanomaterials for science and technology applications due to their intriguing properties like good electrical and thermal conductivities [2, 3], high mechanical stiffness [4], chemical inertness [5] and high aspect ratio [6]. These properties made CNTs to be useful in nano-electronics [7], field emission electron sources [8], high strength composites [9], storage devices [10], sensors [11], bio-medical applications [12] etc. CNTs are considered to be efficient field emitters due to their high aspect ratio, nanoscale tip, good electrical conductivity and high chemical stability. There is a great demand for replacing thermionic emission based conventional electron sources due to high power consumption and emitter deterioration caused by long term heating. Although, CNTs are the potential field emitters in terms of high field emission current and relatively low turn-on and threshold fields, there are some issues, such as poor adhesion and high contact resistance at the CNT-substrate interface, large field screening among adjacent emitters, ion bombardment at the CNT tips, poor emission stability and necessity of high cost and complex equipment for the synthesis and processing, which need to be addressed.

Various approaches have been employed to address the aforementioned issues and optimize the field emission (FE) properties of CNTs using pre/post CNT synthesis treatments. For example; (i) employing an interlayer between the substrate and CNTs, to get good adhesion and reduced contact resistance between CNTs and substrate [13], (ii) catalyst patterned substrates to reduce the field screening among adjacent emitters [14, 15], (iii) nanoparticle coating of a low work function metal/compound/wide band gap material, which may reduce the tunneling barrier for emitting electrons by changing the electronic structure and may also increase the number of emission sites [16-18], (iv) plasma treatment to remove the amorphous carbon present at the tips, modify the tip surface and create more surface electron states by creating defects [19-21], (v) doping through functionalization to reduce the work function, increase the electrical conductivity and emitter site density [22] and (vi) CNT carpet to get highly protruded emitters with high enhancement factor and their electrically good contact with the substrate [23].

Generally, the field emission properties are governed by Fowler-Nordheim theory [24] and field emission current density (J) is expressed as:

$$J = \frac{A\beta^2 E^2}{\phi} exp\left(-\frac{B\phi^{\frac{3}{2}}}{\beta E}\right) \quad (1)$$

where, A (1.544×10⁻⁶ AeVV⁻²), B (6.83×10⁹ eV⁻³ᐟ²Vm⁻¹) are constants, β is a local field enhancement factor at the emitter tip, E is an applied electric field, and ϕ is the work function of a material.

Here, FE current density (J) depends on β and ϕ exponentially, J increases when β increases or/and ϕ decreases. There are certain approaches reported to tune/optimize these two parameters to achieve the superior field emission properties. To increase β, (i) pattern growth of CNT emitters [14, 15]; and (ii) coating of high dielectric constant material along the tubes to avoid field screening between adjacent/neighboring tubes [17] have been discussed. To decrease ϕ, (i) coating of low work function material on CNTs including metals (Cs, Hf) [16, 25], metal-oxides (MgO, ZnO, BaSrO) [18, 26, 27], and other compound materials (LaB$_6$, STO, BN) [15, 17, 28]; and also, (ii) doping/functionalization of (Cs, Li) [29, 30] have been discussed. Majority of these studies monitor the changes in the work function, field enhancement factor and structural aspects to explain the observed field emission properties and have not studied the influence of electronic structure and metal-CNT interface on the field emission. For the first time, Cho *et al.* theoretically explained the modification of electronic structure of CNTs by metal coating through charge transfer and orbital hybridization [31]. Thereafter, a few studies reported influence of electronic structure on field emission properties, which emphasized the role of an increased density of states (DOS) near the Fermi level (0-10 eV) [32-34], decreased work function and improved field enhancement factor [32, 33] and coating induced changes in hybridization from sp$^2$ to sp$^3$-carbon [35, 36]. It is important to investigate the electronic structure and surface-interface properties of metal/metal oxide coated VACNT films since field emission is a surface property which depends not only on work function and density of states near the Fermi level but also on surface hybridization and bonding.

In the present work, we have studied Cu coated VACNTs before and after oxidation to understand the role of metal oxide-CNT interface on field emission properties. Cu is known to get oxidized in ambient conditions and its oxidation state can be tuned by oxidizing it at different temperatures. Hence, it is likely to form different types of interfaces (Cu$_2$O and/or CuO) with CNTs. In a recent work [37], field emission study of CuO decorated amorphous-CNT has been reported. A current density of less than 1 mA/cm$^2$ at a field of 10 V/μm has been obtained. The current density reported in this work is quite low and the role of oxide-CNT interface and electronic structure in tuning the field emission properties has not been addressed. In this work, on subsequent oxidation of Cu (3 nm) coated VACNTs, we were able to achieve a current density of the order of 20 mA/cm$^2$ at a field of 5 V/μm. The results are explained on the basis of critical role played by the metal oxide-VACNT interface in tuning the field emission current density of this system.

## 2. Experimental details

### 2.1 Synthesis of Cu-VACNT films

The VACNT films have been synthesized on Si substrate (n-type: <100>) using thermal chemical vapor deposition technique by double zone horizontal furnace without using any carrier gas. Details of the synthesis process are described in our earlier publications [36, 38]. These grown VACNT films are coated with Cu metal with two thicknesses (3 and 10 nm) using thermal evaporation technique at a base pressure of $5 \times 10^{-5}$ mbar. Subsequently, films were annealed in oxygen atmosphere at 250 °C. Details of all the films used in the present study and their nomenclature are given in Table 1. The nomenclature given in the table is hereafter used in the subsequent text.

**Table 1:** Nomenclature of all deposited CNT films, pristine and metal coated VACNT films and their oxidized counterparts

| S. No. | CNT films | Nomenclature |
|---|---|---|
| 1 | Pristine-VACNT | A0 |
| 2 | Cu(3 nm)/VACNT | A3 |
| 3 | Cu(10 nm)/VACNT | A10 |
| 4 | Oxidized pristine-VACNT | B0 |
| 5 | Oxidized Cu(3 nm)/VACNT | B3 |
| 6 | Oxidized Cu(10 nm)/VACNT | B10 |

*2.2 Analytical characterizations*

Various analytical characterizations have been performed to analyze the morphology, micro-structure, elemental and chemical composition and electronic structure of grown films. The morphology, micro-structure and elemental composition of the films are analyzed using field emission scanning electron microscope (FESEM: FEI Quanta 200F), high resolution transmission electron microscope (HRTEM: JEM2100F) and scanning electron microscope (SEM: TM3000) in energy dispersive mode of X-rays (EDX), respectively. The CNTs have been transferred on to the copper coated TEM grids for HRTEM characterization. The structural aspects of the films are assessed using Raman spectrometer (RENISHAW inVia Raman microscope) with excited laser wavelength of 532 nm. The electronic structure (C 1s, Cu 2p and valence band spectra) of the films is analyzed using X-ray photoelectron spectroscopy (XPS: Omicron nanotechnology, Oxford Instrument Germany) equipped with monochromatic Aluminum source (Al $k_\alpha$: $h\upsilon$=1486.7 eV). The instrument was operated at 15 kV and 20 mA and the pass energy used for recording core level spectra was 20 eV. The work function is estimated for all films using Kelvin probe force microscopy (KPFM: Brukar, Dimension Icon).

*2.3 Field emission studies*

The FE measurements are performed in a high vacuum electrode system, connected in series to turbo molecular and rotary pumps. The base pressure during the measurement is $1\times10^{-7}$ mbar. Here, the CNT film and coated CNTs act as cathode and circular steel plate acts as an anode. The separation between cathode and anode is kept at 300 µm for FE measurements of all films. The range of applied electric field maintained between electrodes is 0-5 V/µm by applying the voltage in steps of 50 V. The high-voltage power supply (Stanford Model: PS350) and an electrometer (Keithley 2000 DMM) are used to apply voltage between electrodes and measure the emitted electron field emission current, respectively.

## 3. Results and discussion

The SEM images of A0 and B3 films are shown in Figs. 1(a) and (b), respectively. Both images clearly show that the nanotubes are vertically aligned. The HRTEM images of A3 and B3 films are shown in figs. 1(c) and (d) showing formation of multi-walled CNTs. The deposition of Cu/oxide coating is clearly seen. The change in contrast observed in figure 1(d) can be attributed to increase in oxide content on oxidation. This is supported by XPS measurements discussed in the subsequent section. Figures 1(e) and 1(f) show the EDX spectra of A0 and A3 films, respectively. The presence of Cu is confirmed by EDX spectrum as shown in figure 1(f).

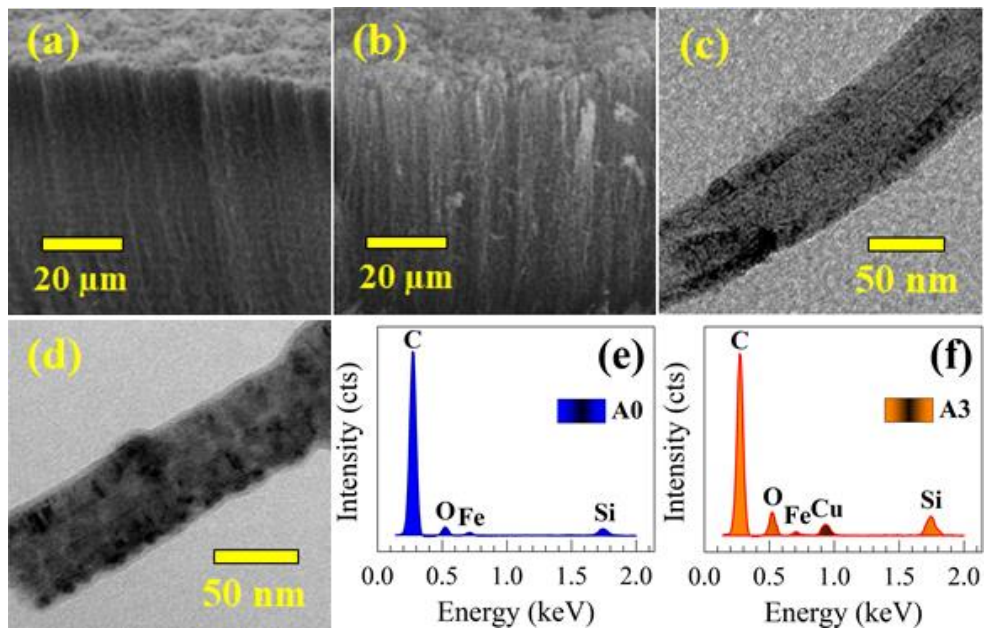

**Fig. 1. SEM images of (a) A0 and (b) B3 films, show vertical alignment of CNTs. HRTEM images of (c) A3 and (d) B3 films showing presence of coating along the nanotube. EDX spectra of (e) A0 and (f) A3 films. A signature of Cu from A3 film confirms the presence of Cu coating on VACNT films.**

Figure 2 shows the Raman spectra of all films, before and after oxidation. Different peaks observed in all films are: (i) the graphitic band (G-band), originating from crystalline hexagonal carbon network, at 1580 cm$^{-1}$; (ii) the defective band (D-band), originating from defective carbon like amorphous carbon, improper hexagon network (heptagon, pentagon etc.), sp$^3$-carbon etc., at 1350 cm$^{-1}$, (iii) 2D-band (2$^{nd}$ harmonic feature of G-band) arising at 2700 cm$^{-1}$ due to the long range order present in hexagonal network, and also (iv) D' and D+D'' bands originate at 1615 cm$^{-1}$ and 2450 cm$^{-1}$ due to defect induced strain in C=C bond along the tube as an asymmetry/shoulder of G-band towards higher wavenumber and overtone mode of LO phonon, respectively. The presence of defects along the tube and strain generated vibrations are other contributing factors to these.

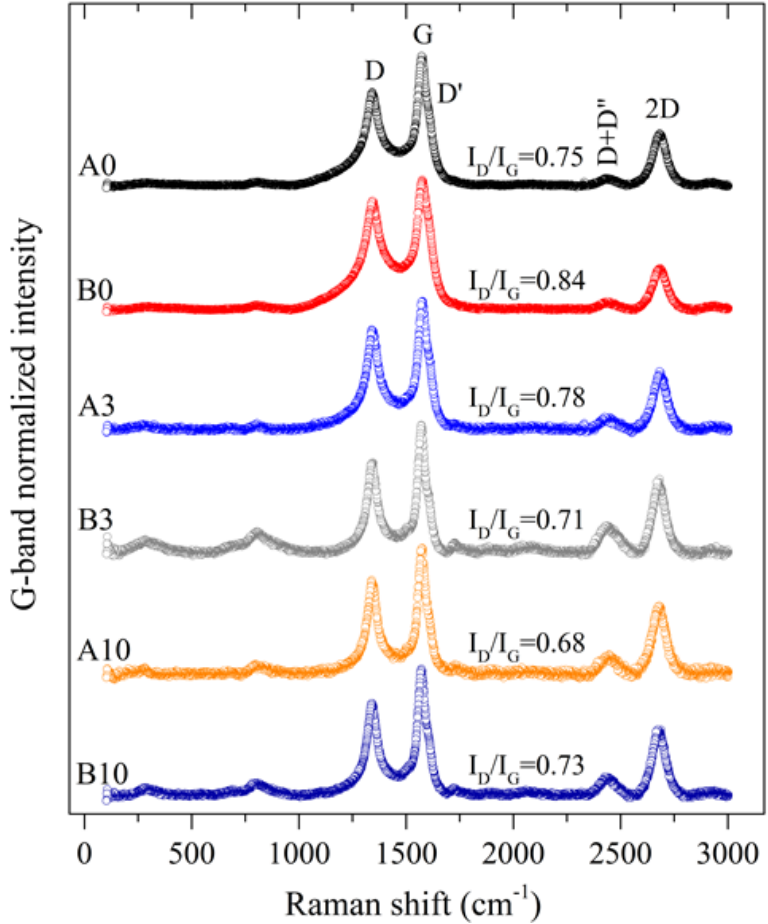

Fig. 2. Raman spectra of CNT films A0, A3, A10, B0, B3 and B10 films. No significant change in $I_D/I_G$ values is observed among pristine (A0), Cu coated (A3 and A10) VACNT films and their oxidized counterparts (B0, B3 and B10).

The peaks observed at the lower wavenumbers (<1000 cm$^{-1}$) are due to various vibrations associated with oxides of Cu [39, 40]. The intensity ratio of D to G peaks ($I_D/I_G$) obtained for different films are shown in Table 2. It is clear that there is no considerable change in $I_D/I_G$ even after Cu coating and their successive oxidation on VACNT films. A little enhancement in intensity of 2D and D+D" bands is seen particularly for B3 film, which is likely due to improvement in the long-range order in the graphitic network and strain due to coating and oxidation.

The KPFM measurements are performed to find the work function of the films through surface potential/contact potential difference. The work function is estimated for all films using the equation [41] given by,

$$V_{CPD} = \frac{(\phi_{Tip} - \phi_{Sample})}{-e} \qquad (2)$$

$$\phi_{Sample} = \phi_{Tip} + eV_{CPD} \qquad (3)$$

where 'e' is the electronic charge, '$V_{CPD}$' is the contact potential difference between CNT film and the metal tip, $\phi_{Tip}$ and $\phi_{Sample}$ are the work functions of the metal tip used and the CNT film, respectively. The work function values are in the range of 4.4-4.8 eV for all films, which suggests that the work function is not affected significantly by Cu coating and is unlikely to play any role in the observed changes in field emission properties.

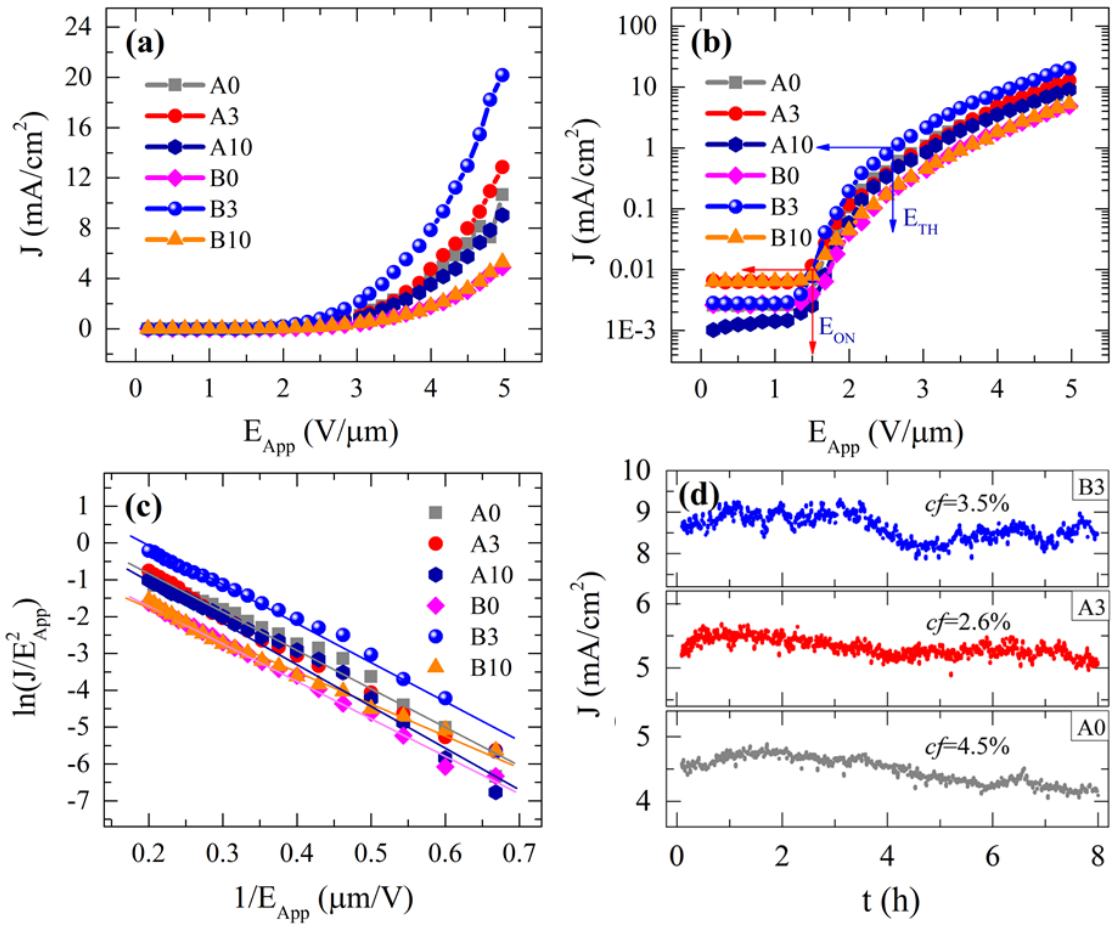

**Fig. 3.** Field emission (a) J-E characteristics, (b) J-E characteristics in semi-log format and (c) corresponding F-N plots of A0, A3, A10, B0, B3 and B10 films. The temporal stability of A0, A3 and B3 films is shown in Fig. (d). The B3 film shows the highest current density (Fig (a); $J_{Max}$=20 mA/cm$^2$). The linearity of F-N plots shows the field emission behavior of the VACNT emitters.

Field emission measurements are performed for all the films in a high vacuum chamber with electrodes in diode geometry. Figure 3 shows the field emission current density versus applied electric field (J-E) characteristics (Fig. 3(a)) of all films, which is also plotted in semi-log format (Fig. 3 (b)) for better clarity. The Fig. 3(c) shows the corresponding Fowler-Nordheim (F-N) plots of the films. FE properties such as turn-on ($E_{ON}$) and threshold ($E_{TH}$) fields, local field enhancement factor (β-factor) and maximum emission current density at maximum applied electric field ($J_{Max}$) are calculated for the films. The turn-on field, $E_{ON}$, defined as the applied electric field required to obtain the emission current density of 10 µA/cm$^2$ from emitters and for threshold field, $E_{TH}$, it is 1 mA/cm$^2$. The maximum current density ($J_{Max}$) is calculated at an applied electric field of $E_{Max}$=5 V/µm for all films. All the calculated FE parameters are presented in Table 2. From Fig. 3(a), it is clear that there is a slight improvement in emission current density after coating of 3 nm thick Cu metal on VACNT film (A3) but current density is significantly enhanced for the same upon its oxidation (B3). Apart from that, other FE parameters including turn-on and threshold fields are also improved for the B3 film. However, both A0 and A10 show decrease in current density upon oxidation.

**Table 2:** Parameters, $I_D/I_G$, work function and all FE parameters of the films

| S. No. | CNT films | $I_D/I_G$ | φ (eV) | $E_{ON}$ (V/µm) | $E_{TH}$ (V/µm) | $J_{Max}$ (mA/cm$^2$) | β-factor |
|---|---|---|---|---|---|---|---|
| 1 | A0 | 0.75 | 4.76 | 1.58 ± 0.2 | 2.98 ± 0.2 | 10.65 ± 0.43 | 6771 |
| 2 | A3 | 0.78 | 4.66 | 1.48 ± 0.2 | 2.98 ± 0.2 | 12.85 ± 0.51 | 7302 |
| 3 | A10 | 0.68 | 4.64 | 1.71 ± 0.2 | 3.13 ± 0.2 | 9.04 ± 0.36 | 5862 |
| 4 | B0 | 0.84 | 4.57 | 1.75 ± 0.2 | 3.55 ± 0.2 | 4.89 ± 0.20 | 6453 |
| 5 | B3 | 0.71 | 4.70 | 1.53 ± 0.2 | 2.62 ± 0.2 | 20.15 ± 0.81 | 6628 |
| 6 | B10 | 0.73 | 4.40 | 1.62 ± 0.2 | 3.55 ± 0.2 | 5.30 ± 0.21 | 7246 |

The local field enhancement factor (β-factor) is calculated for the films by considering the slope of the F-N plot as shown in Fig. 3(c). In general, the β-factor can be expressed as:

$$\beta = -\frac{B\phi^{\frac{3}{2}}}{S_{FN}} \qquad (4)$$

where '$S_{FN}$' is the slope of the F-N plot.

In addition, it is observed that there is no one to one correspondance between $J_{Max}$ and β-factor. From A0 to A3, there is a small enhancement in β-factor, whereas from A3 to B3 the β-factor decreases although $J_{Max}$ is maximum in this case. On the other hand, from A10 to B10, β-factor enhances but the $J_{Max}$ has dropped significantly. Therefore, based on the conventional field emission parameters and structural aspects, which generally explains the FE properties, we could not explain the observed variation in $J_{Max}$ values. Therefore, the correlation between observed variation of $J_{Max}$ and electronic structure of the films was investigated by XPS studies. Field emission temporal stability measurements have been performed for A0, A3 and B3 films at the constant applied electric field of 4.3 V/μm to test the films for electron source device application.

To assess the temporal stability of FE current density, fluctuation in current density (*cf*) is calculated using the following formula [34],

$$cf = \frac{|(x-\bar{x})|}{\bar{x}} \times 100 \qquad (5)$$

where 'x' is the average deviation from the mean current density '$\bar{x}$'. The mean current density and *cf* values are found to be 4.5 mA/cm$^2$, 5.3 mA/cm$^2$, 8.6 mA/cm$^2$ and 4.5%, 2.6%, 3.5%, respectively for A0, A3 and B3 films. Although, there is no much diffrerence in *cf* value of these films B3 film shows good temporal stability at relatively higher field emission current density than A0 and A3 films as shown in Fig. 3(d). This indicates its potential to be used as an electron source in future vacuum microelectronic devices.

The XPS measurements of C 1s, Cu 2p core levels and valence band spectra have been recorded for the films before (denoted as A0, A3 and A10) and after oxidation (denoted as B0, B3 and B10). The Fig. 6 shows C 1s spectra of all films. In our recent study on In coated VACNTs, we had discussed the significance of transformation of sp$^2$- to sp$^3$- carbons in enhancing the current density of the films [36]. Keeping it in mind, each C 1s spectrum is deconvoluted and comprises of five peaks that are ascribed to C=C, C-C, C-O, -O-C=O and a peak related to carbonates [42]. The peak areas are listed in Table 2. Here, C=C and C-C bonds represent $sp^2$- and $sp^3$-hybridized carbons, respectively. It is evident from the table that unlike In/CNT system, the Cu/CNT system does not show any significant variation in fractions of $sp^2$- and $sp^3$-hybridized carbons. Also, $sp^2$ to $sp^3$ ratio has no correlation with the current densities of the films.

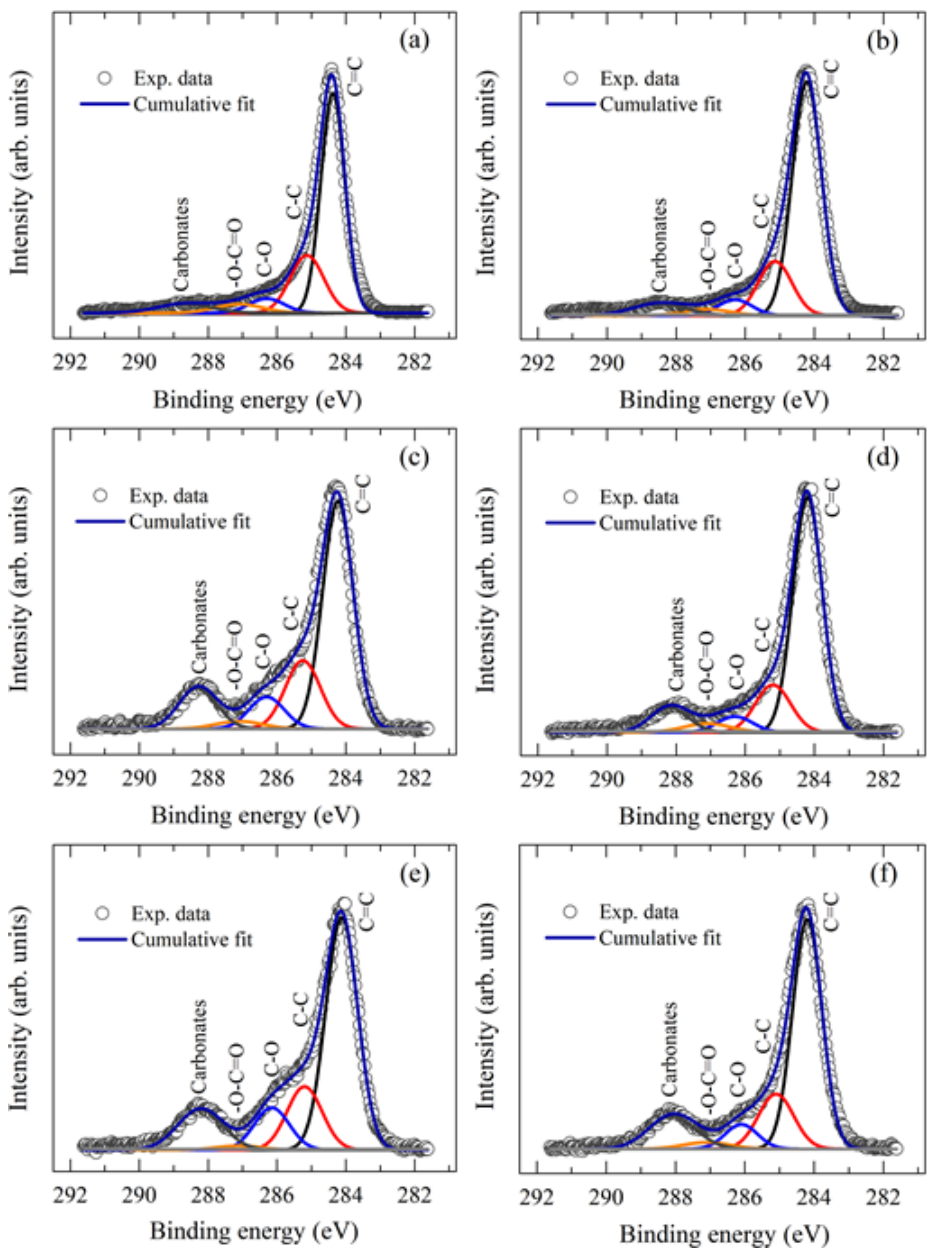

**Fig. 4.** C 1s spectra of (a) A0, (b) B0, (c) A3, (d) B3, (e) A10 and (f) B10 films. The $sp^2/sp^3$ ratio obtained through deconvolution of spectra neither shows any significant change nor has any correspondence with the variation in the current density.

**Table 3:** Parameters of C 1s spectra of A0, B0, A3, B3, A10 and B10 films

| S. No. | CNT films | Area under C=C (sp$^2$) | Area under C-C (sp$^3$) | sp$^2$/sp$^3$ |
|---|---|---|---|---|
| 1 | A0 | 12115 | 4461 | 2.7 |
| 2 | B0 | 12955 | 3458 | 3.8 |
| 3 | A3 | 5929 | 2034 | 2.9 |
| 4 | B3 | 7804 | 1898 | 4.1 |
| 5 | A10 | 6054 | 1803 | 3.4 |
| 6 | B10 | 6262 | 1801 | 3.5 |

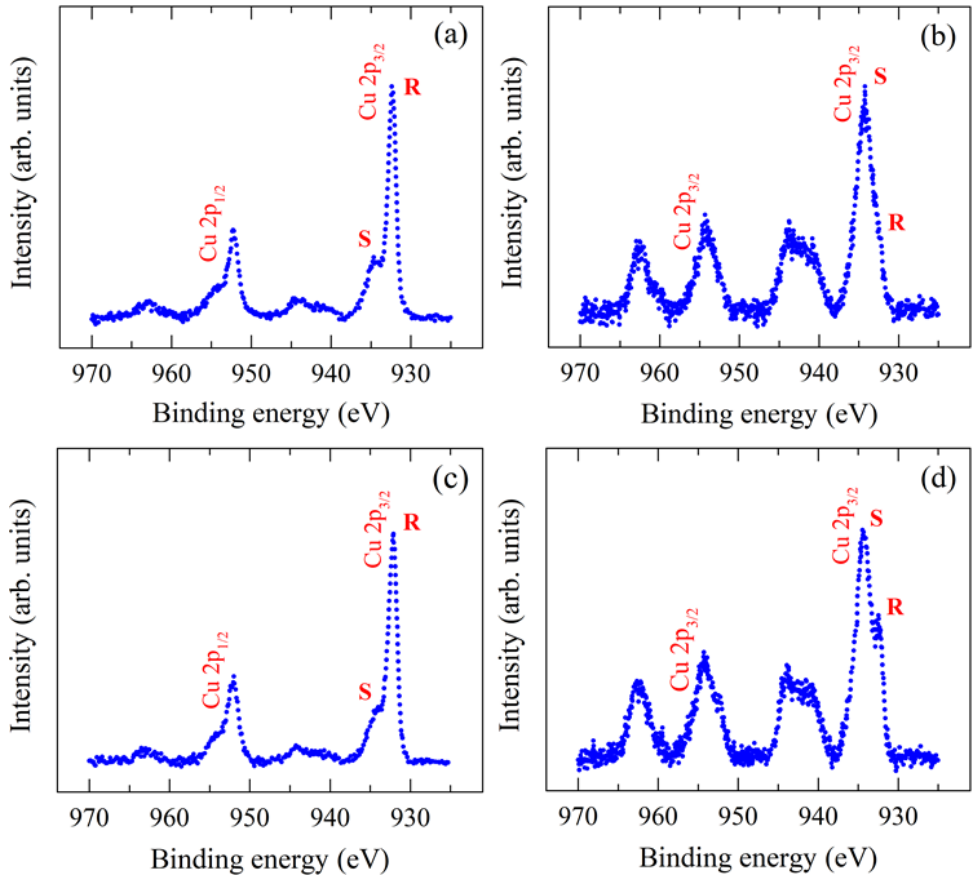

**Fig. 5. Cu 2p spectra of (a) A3, (b) B3, (c) A10 and (d) B10 films. The presence of intense satellite features to Cu 2p$_{3/2}$ and 2p$_{1/2}$ peaks for B3 and B10 films confirms the CuO formation after annealing Cu coated VACNT films in oxygen atmosphere.**

Figure 5 shows Cu 2p spectra recorded for the A3, B3, A10 and B10 films. The main peak R ($2p_{3/2}$) of A3 and A10 films appears around 932.3 eV with a shoulder (S) at 934.2 eV. On oxidation, the main peak shifts to the higher binding energy and is located around 934.2 eV. The binding energy of the main peak and a clear emergence of the satellite structure at about 9 eV from the main peak in the oxidized films show dominating presence of CuO on CNT surface. The matching of the binding energy values of the main peak shown in figures (b) and (d) with that of the shoulder in the main peak shown in figures (a) and (c), suggests that a small fraction of CuO was present even prior to oxidation. The main peak R for the samples A3 and A10 is attributed to $Cu_2O$. The binding energy of Cu $2p_{3/2}$ peak for Cu metal is also reported to be very close to that of $Cu_2O$. However, we have attributed this peak to the presence of $Cu_2O$ as oxidation of Cu is inevitable since the films were exposed to ambient for a significant period of time. However, presence of a small fraction of Cu cannot be ruled out.

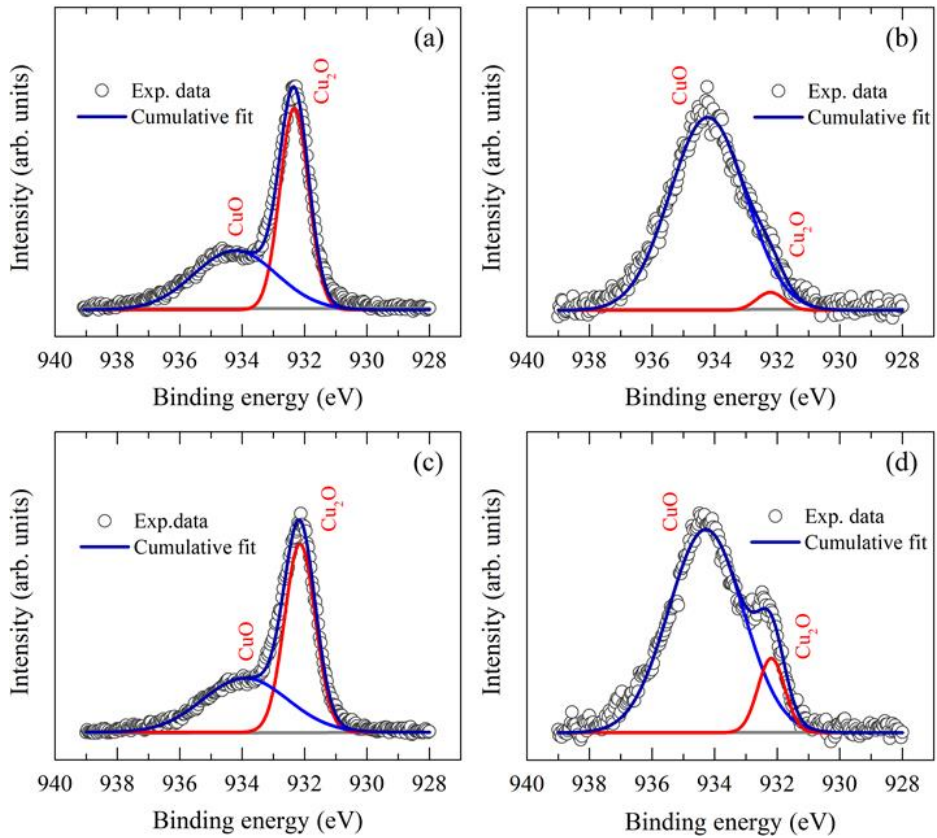

**Fig. 6. Cu $2p_{3/2}$ peak of (a) A3, (b) B3, (c) A10 and (d) B10 films. There is a significant change in CuO/$Cu_2O$ ratio after oxidation of Cu coated VACNT films and this ratio is maximum for the B3 film.**

To find the fraction of $Cu_2O$ and $CuO$ in various films, the main peak was deconvoluted into two peaks as shown in Fig. 6. The area ratio of $CuO/Cu_2O$ of deconvoluted peaks are presented in Table 4. It is to be noted that the ratio of CuO to $Cu_2O$ is maximum for sample B3 for which the highest current density is obtained. The current density gets reduced as the fraction of $Cu_2O$ increases. From the above discussion, it is clear that prior to oxidation $Cu_2O$ is predominantly present on the surface of CNTs with some fraction of CuO. On oxidation, significant fraction of $Cu_2O$ gets converted into CuO modifying the interface between CNT and oxidized Cu species.

**Table 4:** Area ratio of deconvoluted peaks ascribed to CuO and $Cu_2O$ of A3, B3, A10 and B10 films

| S. No. | CNT films | Area ratio of $CuO/Cu_2O$ |
|---|---|---|
| 1 | A3 | 0.8 |
| 2 | A10 | 0.8 |
| 3 | B3 | 29.2 |
| 4 | B10 | 7.6 |

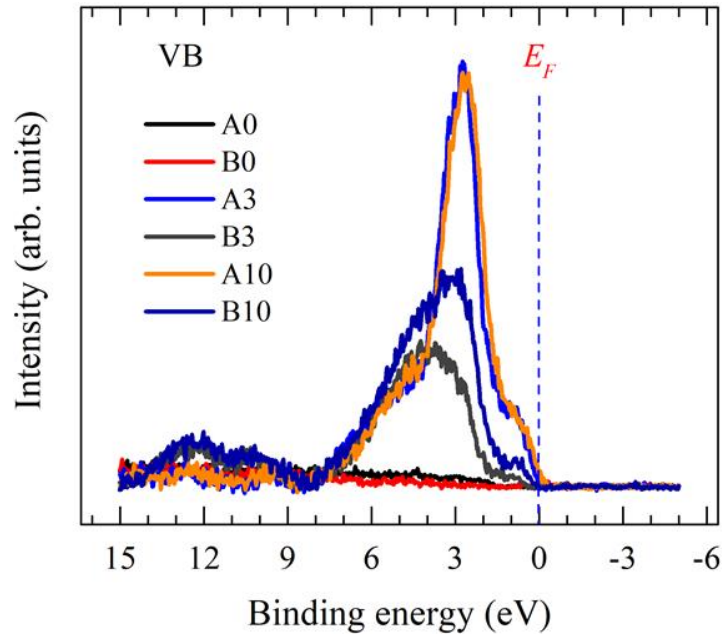

**Fig. 7. Valence band spectra of (a) A0, (b) B0, (c) A3, (d) B3, (e) A10 and (f) B10 films.**

Figure 7 shows the valence band spectra of the films prior to and after oxidation. As shown in the figure, the spectra corresponding to pristine VACNTs show reduced DOS close to the Fermi level even after oxidation. One can distinguish two energy regions in the valence band spectra of Cu coated VACNTs. One region (from 5 eV up to the Fermi level) is dominated by Cu 3d states, pure Cu 3d between 4 and 2 eV and hybridized with Cu 4s and O 2p near the Fermi level [43]. The other region (5-8 eV) has a more pronounced O 2p character [44, 45]. The spectra of Cu coated VACNTs prior to oxidation resemble valence band spectra reported for $Cu_2O$ [46]. These show an appreciable increase in the DOS close to the Fermi level. A peak around 2.7 eV corresponds to Cu 3d states. The hump around 5 eV suggests that apart from $Cu_2O$, a small fraction of CuO is also present. This is in agreement with the analysis of Cu 2p core level spectra of these films, discussed above. An asymmetric feature located around 1.2 eV just below the Fermi level may be attributed to Cu 4sp states, which suggests presence of a small fraction of Cu [46]. However, the spectra do not distinctly show DOS crossing the Fermi level for films A3 and A10, clearly suggesting that Cu, even if it is there, is present at a very low concentration. On oxidation, the DOS near the Fermi level get reduced and the peak gets broadened with spectral weight shifting towards higher binding energy side. The broadening is due to hybridization of Cu 3d and O 2p bands and indicates the formation of CuO. However, it is to be noted that the peak position of the broadened peak in the case of B10 is quite close to that of A3 and A10. This suggests that the fraction of $Cu_2O$ in B10 is more than that of B3. It is in agreement with our analysis of Cu 2p core level spectra. From the aforementioned analysis it is evident that Cu coated on the VACNTs is predominantly oxidized and formation of $Cu_2O$ takes place at the interface of CuO and VACNTs. $Cu_2O$ is expected to have a full 3d shell ($3d^{10}$). The electrons located in the narrow 3d shells are localized whereas CuO has an open shell structure wherein Cu 3d-O 2p shells are hybridized. As a consequence, the electrons residing in the hybridized shells are more delocalized/itinerant. The itinerant electrons participate more readily in the tunneling process resulting into better current density. In the present case, the value of current density has a one to one correspondence with CuO to $Cu_2O$ ratio. Presence of larger fraction of $Cu_2O$ at the interface results in reduced values of current density and vice versa.

The role of interface in the tunneling process is shown in Fig. 8 with the help of an energy band diagrams of VACNT/CuO and VACNT/$Cu_2O$. CNT is a good conducting 1D nanostructure having work function of ~4.8 eV. Both, CuO and $Cu_2O$ are p-type semiconductors having work functions and electronic band gap values of ~5.31 eV, ~5.27 eV and ~1.4 eV, ~2.17 eV, respectively [45, 47]. It is obvious that the formation of Schottky junction between CNT and CuO/Cu$_2$O as CNT is a highly conducting material whereas CuO and Cu$_2$O are semiconducting materials. A downward band bending takes place at both the junctions/interfaces, VACNT/CuO and VACNT/Cu$_2$O, as electrons migrate from low work function material (VACNT) to high work function materials (CuO and Cu$_2$O) as shown in the Fig. 8.

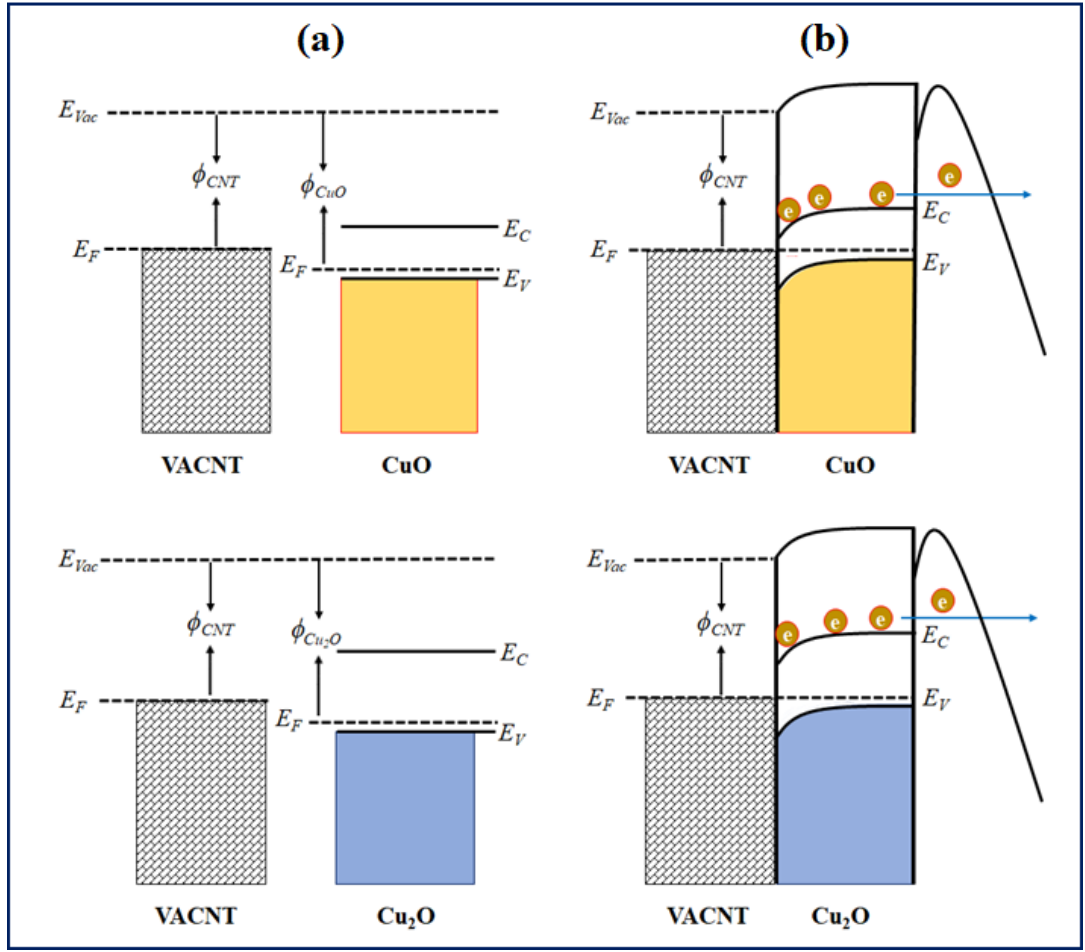

**Fig. 8. Energy band diagrams of VACNT/CuO and VACNT/Cu$_2$O systems; (a) before and (b) after contact.**

From Fig. 8(b), it is clear that, due to the difference in bandgap, there is a large barrier for electron transportation at the interface of VACNT/Cu$_2$O as compared to VACNT/CuO, which in turn reduces/affects the field emission properties of the VACNT/Cu$_2$O films. The barrier height at VACNT/Cu$_2$O interface (A3) is almost 1 eV higher than that of VACNT/CuO interface, which has resulted in more than 150% enhancement of J$_{Max}$.

**Table 5:** Field emission current densities of various metal-oxide coated CNT systems

| S. No. | Metal-oxide coated CNT films | J$_{Max}$ (mA/cm$^2$) | References |
|---|---|---|---|
| 1 | CNT/SiO$_2$ | 0.1 | [18] |
| 2 | CNT/MgO | 0.1 | [18] |
| 3 | CNT/NiO | 2.0 | [48] |
| 4 | CNT/In$_2$O$_3$ | 3.6 | [49] |
| 5 | CNT/IrO$_2$ | 1.0 | [50] |
| 6 | CNT/RuO$_2$ | 4.0 | [51] |
| 7 | CNT/SrTiO$_3$ | 10.3 | [17] |
| 8 | CNT/ZnO | 0.6 | [52] |
| **9** | **CNT/CuO** | **20.1** | **Present work** |

In Table 5, the field emission current densities of various metal oxide coated CNTs as reported in the literature are listed. It shows that the B3 film has the highest J$_{Max}$ value as compare to the reported metal-oxide coated CNTs. So, this oxidized Cu metal coated CNT can be a potential field emitter to replace conventional electron sources in vacuum microelectronic devices upon further optimization.

## 4. Conclusions

The present study reports field emission properties of Cu coated VACNTs. The current density was found to be the maximum for 3 nm thick Cu coated VACNTs after oxidation. The study highlights the importance of studying the electronic structure in order to understand and optimize the field emission properties. A critical analysis of the interface between Cu species and VACNTs through photoemission studies has led us to the importance of presence of Cu$_2$O at the interface of CuO and VACNTs, which in turn controls the current density of these films. The improved field emission properties of this film (3 nm thick Cu coated VACNTs after oxidation) are attributed to the presence of CuO at VACNTs which has itinerant electrons near the Fermi level due to its open shell structure and also having a small barrier at the CuO-CNT interface unlike in the case of $Cu_2O$ for unoxidized counterparts which has a closed shell structure with relatively large barrier at its interface with CNT.

## Acknowledgements

The experimental facilities at UFO Laboratory FIST (DST, Government of India) in the Department of Physics and CRF, IIT Delhi are greatly acknowledged. We thank Materials Research Centre, MNIT Jaipur for providing the XPS facility.